\documentclass[twocolumn,twocolappendix]{aastex631}

\usepackage{graphicx}
\usepackage{gensymb}
\usepackage{float}
\usepackage{subfigure}
\usepackage{verbatim}
\usepackage{makecell}
\usepackage{multirow}
\usepackage{appendix}

\begin{document}

\title{A Tale of Two Type IIb Supernovae: Radio-loud and Radio-silent Candidate Counterparts to Sub-solar Mass Gravitational-wave Triggers}

\author[0009-0007-1842-7028]{Tanner O'Dwyer}
\affiliation{William H. Miller III Department of Physics and Astronomy, Johns Hopkins University, Baltimore, Maryland 21218, USA.}
\correspondingauthor{todwyer1@jh.edu}
\author[0000-0001-8104-3536]{Alessandra~Corsi}
\affiliation{William H. Miller III Department of Physics and Astronomy, Johns Hopkins University, Baltimore, Maryland 21218, USA.}
\author[0009-0004-7268-7283]{Raphael~Baer-Way}
\affiliation{National Radio Astronomy Observatory, 520 Edgemont Road, Charlottesville, VA 22903, USA.}
\author[0000-0002-5619-4938]{Mansi~M.-Kasliwal}
\affiliation{Division of Physics, Mathematics and Astronomy, California Institute of Technology, Pasadena, CA 91125, USA}
\author[0000-0002-0844-6563]{Poonam~Chandra}
\affiliation{National Radio Astronomy Observatory, 520 Edgemont Road, Charlottesville, VA 22903, USA.}
\author[0000-0001-6295-2881]{David L.~Kaplan}
\affiliation{Center for Gravitation, Cosmology, and Astrophysics,
Department of Physics \& Astronomy, University of Wisconsin-Milwaukee, Milwaukee, WI, USA}
\author[0000-0002-2557-5180]{Kunal~P.~Mooley}
\affiliation{Indian Institute of Technology Kanpur, Kanpur 208016, UP, India}
\author[0009-0004-7280-1357]{Deepika Yadav}
\affiliation{Indian Institute of Technology Kanpur, Kanpur 208016, UP, India}

\begin{abstract}
On 2025 November 12, the LIGO-Virgo-KAGRA collaboration reported S251112cm, a gravitational-wave (GW) candidate from a compact binary merger with at least one sub-solar-mass component. Optical follow-up identified SN\,2025adtq, a Type IIb supernova (SN) within the localization volume that exploded a few days before the merger. SN\,2025adtq is the second Type IIb SN found to be potentially coincident with a sub-solar-mass GW candidate, after SN\,2025ulz (S250818k). A key question is whether such SNe harbor fast, possibly collimated ejecta from a central engine, as invoked in ``superkilonova'' scenarios for forming sub-solar-mass neutron stars. We present deep Karl G. Jansky Very Large Array observations of SN\,2025adtq from the JAGWAR program, spanning 139--216 days after the GW trigger at 6--10\,GHz. The radio properties of SN\,2025adtq  differ substantially from those of SN\,2025ulz: whereas SN\,2025ulz has a significant 6--10\,GHz counterpart, SN\,2025adtq remains undetected to $3\sigma$ limits of $\approx7.8$\,$\mu$Jy, implying a radio luminosity density one order of magnitude below that of SN\,2025ulz at a comparable epoch, excluding jet models that remained viable for SN\,2025ulz. This diversity in radio properties calls for systematic multi-wavelength follow up observations of candidate sub-solar events in future runs of the LIGO-Virgo-KAGRA detectors. 
\end{abstract}


\section{Introduction} \label{sec:intro}
On 2025 August 18, the LIGO–Virgo–KAGRA \citep[LVK;][]{2015CQGra..32g4001L,2015CQGra..32b4001A,2021PTEP.2021eA101A} collaboration reported S250818k, a subthreshold gravitational-wave (GW) candidate consistent with a binary neutron star (NS) merger potentially involving a subsolar-mass NS with an astrophysical probability of \(p_{\rm astro}\approx29\%\), with a terrestrial origin favored at $\approx$71\% \citep{2025GCN.41437....1L, 2025GCN.41440....1L}. Optical follow-up identified a young stripped-envelope Type IIb supernova (SN) dubbed SN\,2025ulz, temporally and positionally compatible with the GW trigger within the large localization errors \citep{Kasliwal_2025, 2026A&A...714A.284A, 2025arXiv251024620X, 2026ApJ..1001L..20H}. \citet{2026A&A...714A.284A} estimated a $\sim9\%$ chance association between SN\,2025ulz and the candidate GW event S250818k, as well as a $73.5\%$ chance that SN\,2025ulz and S250818k are unrelated, when factoring in the fact that the GW event had a relatively low $p_{\rm astro}$. Radio observations by \citet{O’Dwyer_2026} identified a radio counterpart to SN\,2025ulz, consistent with the commonly observed non-thermal emission from SN ejecta interacting with circumstellar material. However, radio emission powered by an off-axis jet peaking at $\sim 50-100$\,days after the GW trigger remains compatible with the radio data.

Less than three months after S250818k, on 2025 November 12 at 15:18:45 UTC, a second sub-solar-mass candidate dubbed S251112cm 
\citep[false alarm rate (FAR) of $\approx 1/4$\,yr;][]{LVK2025_GCN42690}, was identified in LIGO Hanford, LIGO Livingston, and Virgo data. Contingent on being astrophysical, S251112cm had a $>99\%$ probability of harboring a compact object lighter than $1\,M_\odot$
\citep{LVK2025_GCN42650}. The LVK public alerts reported a 90\% credible GW sky localization area of $\approx 1681\,\mathrm{deg}^2$ \citep[containing more than 9000 cataloged galaxies;][]{2025GCN.41440....1L,Cook2025_GCN42653,Cook2025_GCN42693}, and a GW luminosity distance of $93\pm27$\,Mpc \citep{2025GCN.41440....1L,2025GCN.42650....1L,2025GCN.42690....1L}. Although the LVK collaboration did not report a specific \(p_{\rm astro}\), its false-alarm-rate of about one per 4-6 years suggests \(p_{\rm astro}\sim50\%\) for S251112cm, assuming a FAR–\(p_{\rm astro}\) relationship similar to that of typical BNS searches \citep[see ][Section~4.1 footnote 9]{2026A&A...714A.284A}. 

\citet{Hall2026} identified SN\,2025adtq within the sky localization area of S251112cm, utilizing a search time window that extended to 4\,d before the merger. They also estimated a 2\%--9\% probability that an unrelated Type IIb SN could have exploded within the corresponding volume during the 4\,d window, under the assumption that S251112cm is astrophysical in origin. Once the presence of a coincident Type IIb SN is taken as given, the odds favor chance alignment, so the association remains suggestive rather than established. 

The potential association of sub-solar-mass GW candidates with core-collapse SNe has renewed the community's interest in  previously proposed
``superkilonova'' models, invoking the formation of $r$-process elements in collapsars \citep{2019Natur.569..241S,2022ApJ...941..100S}. The term ``superkilonova'', in particular, has been extended to include mergers of sub-solar NSs formed within a collapsing core via accretion-disk fragmentation or core fission induced by extreme rotation
\citep{1998AstL...24..206I,2007ApJ...658.1173P,2016MNRAS.463.1642P,2002ApJ...579L..63D,2024ApJ...971L..34M,2022ApJ...941..100S,2025ApJ...991L..22C}, ultimately producing a  kilonova inside a supernova 
\citep{2007ApJ...658.1173P,2024ApJ...971L..34M,2025ApJChenMetzger,Kasliwal_2025,2025arXiv250521617L}. 

While the superkilonova scenario may accommodate associations between sub-solar-mass GW triggers and core-collapse SNe from rapidly rotating progenitors, it remains largely speculative: it has yet to be confirmed observationally and faces several theoretical challenges \citep{2026ApJ..1001L..20H,2025arXiv251024620X,2025arXiv251023728O,2026ApJ...996L..24Y}. Indeed, stellar evolution models generally predict a NS mass of
$\sim1.2\,M_\odot$ \citep{2025PhRvL.134g1403M}, and the lightest NSs reported
observationally have masses of $1.02\pm0.17\,M_\odot$
\citep{2015A&A...577A.130F,2016ARA&A..54..401O}. On the other hand, stable NSs with masses down
to $\sim0.1\,M_\odot$ may exist in highly neutron-rich
environments \citep{2002A&A...385..301H}. GW observations may provide a direct means of identifying such events.\citep[e.g.,][and references therein]{2019PhRvL.123p1102A,2026arXiv260723119C,2026arXiv260925225B,2026arXiv260811387T}.

As highlighted by \citet{Hall2026}, supporting EM evidence for the association between S251112cm and SN\,2025adtq would be $r$-process signatures or an associated gamma-ray burst, which could be revealed by X-ray or late-time radio observations. Indeed, within the superkilonova scenario, the detection of a radio jet afterglow would support a link between SN\,2025adtq and a rapidly rotating progenitor capable of launching mildly to highly relativistic outflows \citep[e.g.,][]{1998Natur.395..663K,Patat_2001,2003ApJ...599..408B,2004Natur.430..648S,2010Natur.463..513S,2006ApJ...639..331G,2013ApJ...778...63H,2013ApJ...778...18M,2016ApJ...830...42C,Alexander_2017,2017ApJ...847...54C,2018ApJ...863...32D,Dobie_2018,Mooley_2018,2020ApJ...902...86H,2020ApJ...893..132H,Makhathini_2021}. However, as discussed in \citet{O’Dwyer_2026} for SN\,2025ulz, radio emission can also arise from the interaction of the SN ejecta with circumstellar material (CSM). Because this mechanism is commonly observed in Type IIb and Ib/c SNe \citep[e.g.,][]{1982ApJ...259..302C,Chevalier_1998,1998ApJ...509..861F,2003ApJ...592..900S,2004MNRAS.349.1093R,2007ApJ...671.1959W,Chevalier_2010,2018SSRv..214...27C,2016ApJ...818..111K,2021ApJ...923L..24S,2024MNRAS.534.3853R,Chandra2025,2025PASA...42...50S,2003ApJ...599..408B,2006ApJ...638..930S,2014ApJ...782...42C,2016ApJ...830...42C,2023ApJ...953..179C,2025arXiv251208822O}, extensive radio observations are needed to test whether the emission is more likely due to slow ejecta interacting with a dense CSM or to faster ejecta with weaker CSM interaction.  

Motivated by the above considerations, here we present deep radio observations of the SN\,2025adtq field carried out with the Karl G.\ Jansky Very Large Array (VLA) through our ``Jansky VLA mapping of Gravitational Waves as Afterglows in Radio'' (JAGWAR) program (VLA/22B-235; PI: Corsi). Our paper is organized as follows.  In Section \ref{sec:follow}, we summarize the multi-wavelength follow up observations of S251112cm. In Section \ref{sec:obs}, we present our radio observations of SN\,2025adtq. In Section~\ref{sec:model} we describe the constraints that our radio observations of SN\,2025adtq place on both the SN--CSM interaction model and an off-axis GRB afterglow scenario, and in Section~\ref{sec:conclusion} we summarize and conclude.

\begin{figure}
\centering
\includegraphics[width=0.45\textwidth]{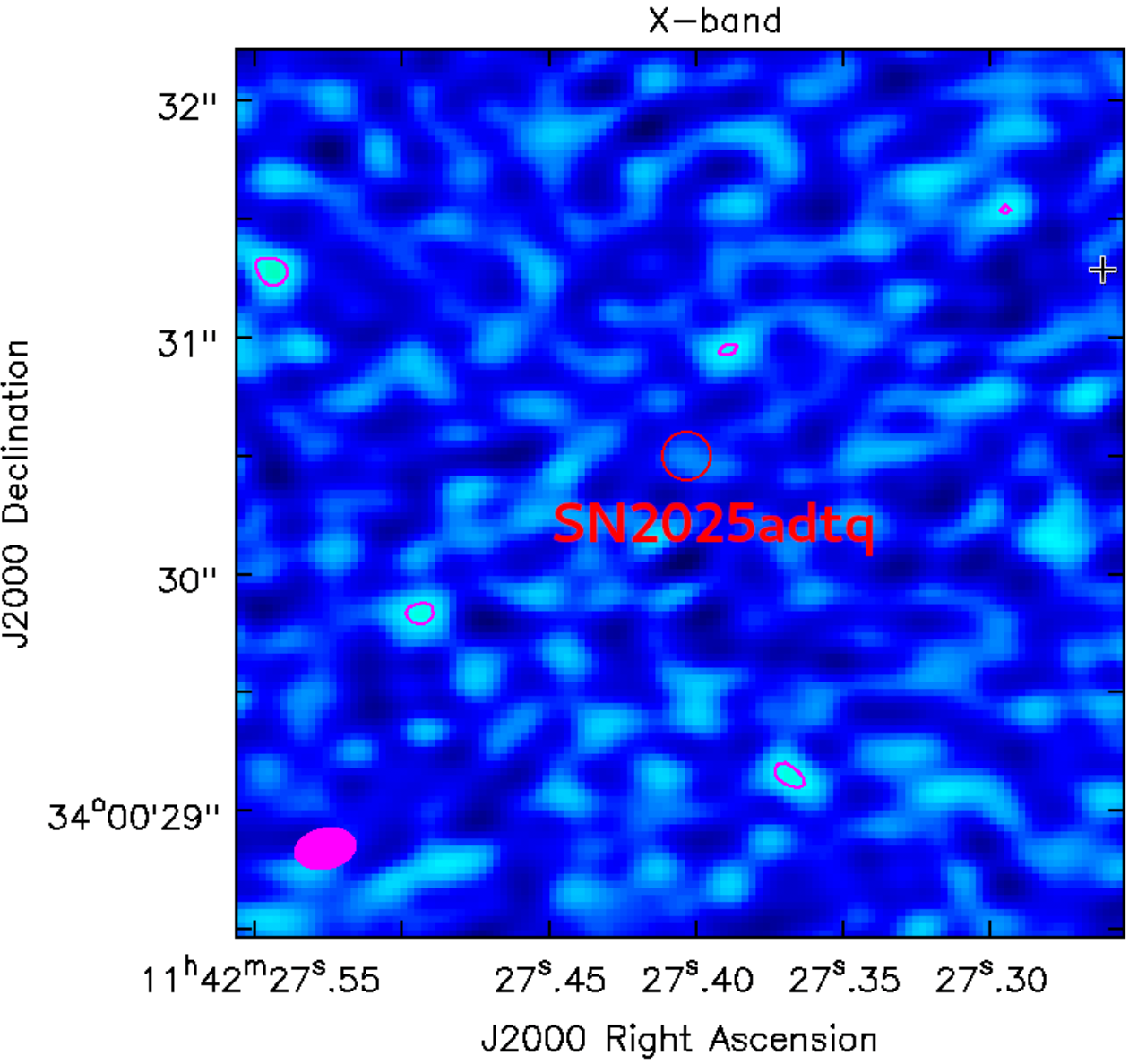}
\caption{Co-added Observations of the SN\,2025adtq field spanning 139-216 days post explosion carried out at $\nu\approx 10$\,GHz under our JAGWAR program with the VLA in its A configuration. The magenta circle marks a region of radius equal to the nominal FWHM of the VLA synthesized beam in A configuration. The actual beam is 
$0.26\arcsec \times 0.17\arcsec$ at PA $-79.0^\circ$
for X band, centered on the position of SN\,2025adtq. In Table \ref{tab:Radio data}, we report the $3\sigma$ upper limit radio flux densities measured within this circular region. The actual VLA beam is plotted as a magenta ellipse. The contour lines show flux levels of 3$\times$RMS.}
\label{fig:Co-add}
\end{figure}

\section{Multi-wavelength follow up of S251112\lowercase{cm}}
\label{sec:follow}
Despite targeted searches were conducted over a large portion of the LVK localization region of S251112cm, no high energy counterpart to it was identified. \textit{Fermi}-GBM was observing 68.6\%
of the localization probability of S251112cm at the event time and set $3\sigma$ flux upper
limits of $(1.0$--$7.7)\times10^{-7}\,\mathrm{erg\,cm^{-2}\,s^{-1}}$
(10--1000\,keV) on 0.128--8.192\,s timescales, corresponding to luminosities
$\lesssim(0.2$--$2)\times10^{48}\,\mathrm{erg\,s^{-1}}$ at the distance of S251112cm 
\citep{Bala2025_GCN42655}; Glowbug reported comparable limits of
$(0.2$--$6.0)\times10^{-7}\,\mathrm{erg\,cm^{-2}\,s^{-1}}$ (50--2000\,keV)
over a complementary part of the sky \citep{Kerr2025_GCN42746}. MAXI/GSC
scanned 72\% of the 90\% region within one orbit and found no new source above
a typical $1\sigma$ limit of 20\,mCrab (2--20\,keV;
\citealt{Hiramatsu2025_GCN42659}). Galaxy-targeted X-ray observations with
\textit{Swift}/XRT (105 pointings, $12.7\,\mathrm{deg}^2$;
\citealt{Page2025_GCN42681}) and EP-FXT (100 pointings, $88\,\mathrm{deg}^2$;
\citealt{Wu2025_GCN42714}) yielded only cataloged sources or faint uncataloged
ones, with 0.5--10\,keV fluxes of
$\sim10^{-13}$--$10^{-12}\,\mathrm{erg\,cm^{-2}\,s^{-1}}$ in the Einstein Probe (EP)-FXT case,
none of which stood out as a likely counterpart.

In the optical, wide-field searches began within an hour of the trigger and
together tiled most of the GW localization region of S251112cm \citep{Ackley2025_GCN42658,Gassert2025_GCN42674,Gillanders2025_GCN42666,Hello2025_GCN42699,Smith2025_GCN42682,Groot2025_GCN42663,Anand2025_GCN42677,Liu2025_GCN42722,Hall2025_GCN42691,Jeong2025_GCN42867,Smith2025_GCN42682,Taguchi2025_GCN42725}). The deepest
coverage came from the Vera C.\ Rubin Observatory, which imaged
$\approx850\,\mathrm{deg}^2$ (39\% of the probability) in $g$ and $i$ on
several nights between November 14 and 22, to median depths of
$g\approx24.2$--24.8 and $i\approx23.2$--23.7\,mag
\citep{MacBride2025_GCN42707,Anand2025_GCN43257}. 

The Multi-messenger Tool for Rapid Object Vetting and Examination (TROVE) was used to search within the 95th percentile localization region of S251112cm for candidate optical counterparts in host galaxies at the approximate luminosity distance of S251112cm. Of the four candidates that survived the TROVE vetting of all sources reported to the Transient Name Server (AT\,2025adht, AT\,2025adgp, AT\,2025adhf, AT\,2025adhs;
\citealt{Franz2025_GCN42675}), AT\,2025adhf was classified as a Type Ia
supernova at $z=0.098$, as were AT\,2025adim ($z=0.091$) and AT\,2025adiw
($z=0.165$), all beyond the GW distance
\citep{Santos2025_GCN42724,Bommireddy2025_GCN42796}. AT\,2025adht faded by only
$\approx0.2$\,mag in $r$ between 2.4 and 10.2\,days after the merger
\citep{Burkhonov2025_GCN42698,Gillanders2025_GCN42825} and lies in a host with
a photometric redshift of $0.14\pm0.03$, pointing to a background SN
near peak rather than a fast-evolving, kilonova-like transient. 

A Zwicky Transient Fcility (ZTF) search deliberately
retained candidates consistent with young stripped-envelope SNe \citep{Anand2025_GCN42677}. Additional unclassified candidates with nearby
host redshifts were reported by the 2.5-meter Wide Field Survey Telescope \citep[WFST;][]{Liu2025_GCN42722} and the Dark Energy Camera on the 4-meter Blanco telescope \citep[DECam;][]{Hall2025_GCN42691}, with later photometry of several candidates from the Space Variable Objects Monitor \citep[SVOM;][]{Ma2025_GCN42777}. In the analysis of the DECam, ZTF, and Wendelstein
data (which together covered 56\% of the localization probability within 48\,hr
of the merger, or $\approx60\%$ including WFST), \citet{Hall2026} find no
kilonova candidate, and ruled out a canonical kilonova with a probability of $\approx29\%$. An independent campaign that vetted 456 candidates likewise
found no likely counterpart \citep{Vieira2026}. However, \citet{Hall2026} extended their analysis to transients appearing up to 4 days before the merger, motivated by the ``superkilonova'' disk-fragmentation model, in which a rapidly rotating massive star collapses, its disk fragments into small NSs that later merge, and a stripped-envelope SN precedes the GW signal by up to a few days. From this analysis, SN\,2025adtq emerged as an interesting candidate: a spectroscopically confirmed Type IIb inside the 3D GW localization located at about 140\,Mpc, that exploded about 2 days before the merger.

SN\,2025adtq is located at $\alpha=11^{\rm h}42^{\rm m}27\fs40$, $\delta=+34^\circ00'30\farcs5$\footnote{\url{https://www.wis-tns.org/object/2025adtq}}(J2000), which lies in the northern lobe of the localization recovered in the WFST difference images \citep{Liu2025_GCN42722}. The WFST measurement,
$r=21.23\pm0.11$\,mag on 2025 November 11.94 UT, is the earliest detection of
the source and precedes the merger by 0.70\,d. ZTF detected it at
$r=20.87\pm0.15$\,mag 0.85\,d after the merger, and Pan-STARRS reported it to
the Transient Name Server (TNS) from an $i=20.22\pm0.09$\,mag detection at
$+2.96$\,d \citep{Chambers2025_TNSTR}; it had brightened to $g\approx19.1$\,mag by $+10.9$\,d. The SN is offset by $8\farcs7$ from a galaxy at $z=0.033$ (140\,Mpc), marginally consistent with the GW distance range of $93\pm27$\,Mpc. Nebular-phase spectra obtained $\approx100$--120\,d after explosion with HET, the Palomar 200\,inch, and Keck are consistent with a Type IIb SN \citep{Hall2026} at a redshift of \citep[$z=0.0333$;][]{ Rose2026_TNSCR}. Hydrodynamic modeling of the light curve gives an ejecta mass of $\approx2.1\,M_\odot$, a $^{56}$Ni mass of $\approx0.075\,M_\odot$, a kinetic energy of $\approx0.6\times10^{51}$\,erg, and an explosion epoch of MJD\,$60988.7\pm1$, i.e., $2.9\pm1.0$\,d before the merger time of S251112cm. Photometrically the event is a normal Type IIb supernova, unlike the unusually red SN\,2025ulz associated with S250818k. For SN\,2025ulz, \citet{Kasliwal_2025} estimated ejecta mass of $\approx3.0\,M_\odot$, a $^{56}$Ni mass of $\approx0.1\,M_\odot$, a kinetic energy of $\approx1.6\times10^{51}$\,erg. These values are comparable, though systematically higher, than the ones derived for SN\,2025adtq.

\section{VLA Observations and data reduction} \label{sec:obs}

We observed the field of SN\,2025adtq with the VLA in its A configuration at 6 and 10\,GHz. Observations were conducted for $\approx3$ hr per observing band, including calibration. The first epoch began on $\mathrm{MJD}=61131.03$ UT ($\sim$139 days post S251112cm), and the last on 61207.81 UT ($\sim$216 days post-trigger). We used J1147+3501 as phase calibrator and 3C286 for flux density and bandpass calibration. Data calibration and imaging were performed using the automated VLA calibration and imaging pipeline \citep{2020ASPC..527..571K} in \texttt{CASA} \citep{2022PASP..134k4501C}, applying self-calibration solutions to the data. After automated calibration and imaging, all data were also manually inspected for the presence of potential radio-frequency interference (RFI) effects. Manual cleaning was also performed using  interactive \texttt{tclean} in \texttt{CASA} without self-calibration on the first epoch, and results were fully consistent with the automated cleaning and imaging results. 

Using a circular region centered on SN\,2025adtq, with radius equal to the nominal VLA synthesized beam at the configuration and frequency setup of each observation, we found no evidence for emission above $3\times$ the root-mean-square (RMS) of the noise. The last is estimated from each image within a circular region of radius $\approx 10\times$ the nominal FWHM of the synthesized beam centered on the position of SN\,2025adtq ($\alpha=11^{\rm h}42^{\rm m}27\fs40$, $\delta=+34^\circ00'30\farcs5$), using the \texttt{imstat} tool in \texttt{CASA}. We also verified that the noise RMS estimated in this region is consistent with the RMS estimated using a circular region of the same size free of bright sources. Our upper limits are reported in Table \ref{tab:Radio data}. 

Finally, to improve our sensitivity, for each frequency band, we combined all three epochs using the standard measurement set concatenation and imaging procedure in \texttt{CASA}\footnote{\url{https://science.nrao.edu/facilities/vla/data-processing/pipeline/vipl_666}}. This produced a deep continuum image for each band, spanning $139$--$216$ days since the GW trigger. At the location of SN\,2025adtq, we again found no evidence for emission above three times the local image RMS (Figure~\ref{fig:Co-add}).

\begin{figure}
    \centering \includegraphics[width=0.47\textwidth]{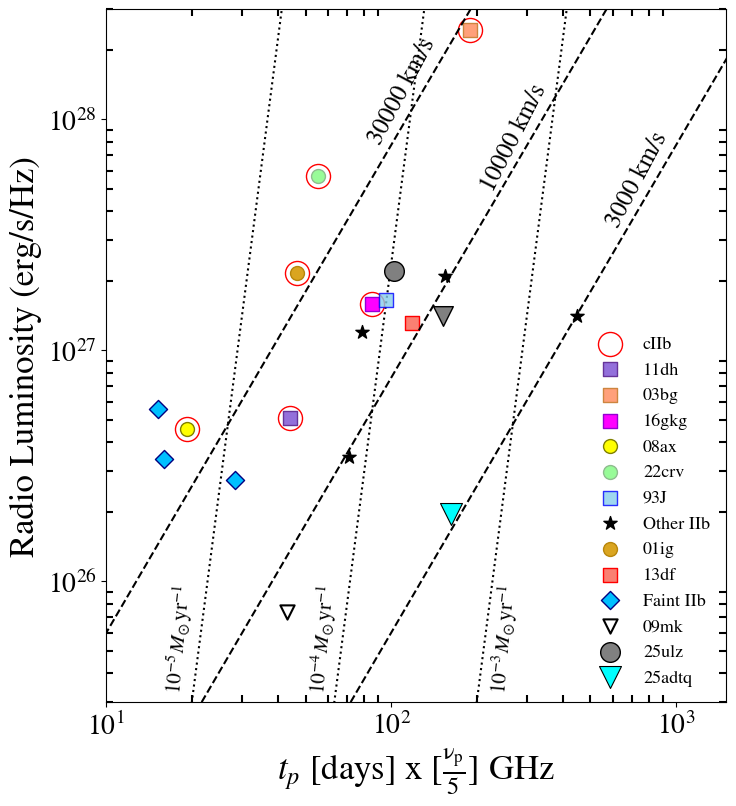}
    \caption{A Chevalier diagram (radio luminosity vs peak time) with the earliest 6\,GHz 3$\sigma$ upper limit of SN\,2025adtq (cyan downward triangle) compared to that of SN\,2025ulz (gray dot detection peak and, downward triangle upper limit) and other Type IIb SNe (as indicated in the legend). We also compare the upper limit to three radio ``faint'' SNe Type IIb (SN\,1996cb,  \citealt{VanDyk1996SN1996cb}; SN\,2008bo,  \citealt{Stockdale2008SN2008bo,Stockdale2008SN2008boRebrightening}; SN\,2012P,  \citealt{Stockdale2012PTF12os}) and two IIb SNe $3\sigma$ upper limits (SN\,2009mk; \citealt{Ryder2010SN2009mk}, SN\,2016gkg, which later brightens see legend; \citealt{Ryder2016SN2016gkg,Nayana2022SN2016gkg}), with historical measurements compiled by \citet{Bietenholz2021RadioSNe}. Lines of constant shock velocity and mass-loss rate are shown. Type cIIb explosions thought to be linked to compact progenitors are highlighted with open red circles.}
    \label{fig:Chevalier}
\end{figure}

\section{Radio modeling} \label{sec:model}
Radio emission from core collapse SNe probes the fastest outflow components, as well as the density and structure of the CSM \citep[e.g.,][and references therein]{1982ApJ...259..302C,Chevalier_1998,1998ApJ...509..861F,1998Natur.395..663K,1998ApJ...497L..17S,1999ApJ...516..788W,2000ApJ...536..195C,2003ApJ...599..408B,2003ApJ...586..356Z,Chevalier_2010,2015MNRAS.450.1430H,2017Sci...358.1579H,2017ApJ...834...28N,2018SSRv..214...27C,2018ApJ...856...56C,2018ApJ...863...32D,2018PhRvL.120x1103L}. In what follows, we describe the constraints that our observations set on the radio emitting ejecta of SN\,2025adtq in the context of two main models: (i)
SN ejecta interacting with the CSM; (ii) a rapidly-rotating progenitor powering a SN plus a fast, collimated ejecta.

\subsection{CSM interaction}
In Figure~\ref{fig:Chevalier}, we show a Chevalier diagram (peak luminosity versus peak time; \citealt{Chevalier_1998}) that places several Type IIb SNe in the context of the synchrotron-self-absorption (SSA) model for radio SNe, where the radio emission is driven by the interaction of the SN shock front with a CSM that is assumed to result from a  progenitor with a constant mass loss. In this Figure we also plot lines of constant SN-shock velocity and mass-loss-rate, following the formulation by  \citet{2012ApJ...752...78S}, where the SN ejecta speed and mass-loss rate of the progenitor can be expressed as follows: 
\begin{equation}
\overline{v}_{\rm sh} \equiv \frac{R_p}{t_p}
\simeq 3.82 \times 10^{5}
L_{\nu_p,26}^{9/19}
\nu_{p,5}^{-1}
\left(\frac{t_p}{1\,\mathrm{day}}\right)^{-1}
\ \mathrm{km\,s^{-1}}.
\label{eq:ssa_velocity}
\end{equation}

\begin{equation}
\dot{M} \simeq 1.95 \times 10^{-8}
L_{\nu_p,26}^{-4/19}
\nu_{p,5}^{2}
\left(\frac{t_p}{1\,\mathrm{day}}\right)^{2}
\left(\frac{v_w}{500\,\mathrm{km\,s^{-1}}}\right)
M_\odot\,\mathrm{yr^{-1}}.
\label{eq:ssa_mass_loss}
\end{equation}

The above relations assume a wind velocity of $v_w=500\,\mathrm{km\,s^{-1}}$, and  make use of the normalizations $L_{\nu_p,26}=L_{\nu_p}/
(10^{26}\,\mathrm{erg\,s^{-1}\,Hz^{-1}})$,
$\nu_{p,5}=\nu_p/(5\,\mathrm{GHz})$, where $t_p$ is the time
since explosion. These relations also assume $p=3$ for the index of the electron energy distribution behind the shock front;
$\epsilon_e=\epsilon_B=0.1$ for the fraction of energy going into accelerating electrons and amplifying magnetic fields; an emitting volume filling
factor $f=0.5$; and negligible free--free absorption.

\citet{2010ApJ...711L..40C} divided Type IIb SNe into two classes,
distinguished mainly by their radio properties. Compact (cIIb) events thought to be associated with Wolf--Rayet-like progenitors with radii $R_\star \sim 10^{11}$\,cm. These
stars lose mass through fast winds ($v_w \sim 1000$\,km\,s$^{-1}$), which
leave a low-density circumstellar medium. Their radio emission evolves
quickly, like that of Type Ib/c events. If the radio peak is set by SSA, their shells move at
$\sim 30{,}000$\,km\,s$^{-1}$, and their wind densities are
$\dot{M}/v_w \approx (0.65$--$13)\times10^{-5}\,M_\odot\,{\rm yr}^{-1}/
(1000\,{\rm km\,s^{-1}})$. Their radio light curves show modulations, which
suggests that the progenitor's mass loss varied over time. Extended (eIIb)
events are thought to originate from supergiants with
$R_\star \sim 4\times10^{13}$\,cm. These stars sit inside slow, dense winds:
for $v_w = 10$\,km\,s$^{-1}$,
$\dot{M}/v_w \approx 4\times10^{-5}\,M_\odot\,{\rm yr}^{-1}/
(10\,{\rm km\,s^{-1}})$. While the mass-loss rate is similar to those of cIIb
events, the wind is 100 times slower, so its density is substantially higher. Extended events therefore evolve slowly in radio, and their
shells are slower, $\sim 10{,}000$\,km\,s$^{-1}$.

In Figure~\ref{fig:Chevalier}, we compare the earliest $3\sigma$ upper limit for SN\,2025adtq at 6\,GHz (cyan downward-pointing triangle) with the 10\,GHz detection of SN\,2025ulz (gray filled circle), likely near its radio peak, and radio measurements of other Type IIb SNe. The SN\,2025adtq limit corresponds to $L_\nu\lesssim2\times10^{26}\,\mathrm{erg\,s^{-1}\,Hz^{-1}}$, a luminosity level below that of SN\,2025ulz and many of the plotted compact (cIIb) linked to Wolf–Rayet stars or stripped binaries, and extended (eIIb) events which arise from extended supergiants. For completeness, we also include in this plot other SNe on the faint end of the observed Type IIb radio population. Specifically, SN\,1996cb \citep{VanDyk1996SN1996cb}, which was placed in the cIIb class by \citet{Chevalier_2010} on the basis of its early light curve and radio properties, although its hydrogen envelope mass remains uncertain. Another radio faint Type IIb is  SN\,2008bo \citep{Stockdale2008SN2008bo,Stockdale2008SN2008boRebrightening}, which was placed in the cIIb class by \citet{Chevalier_2010} because its optical spectra resemble those of SN~2008ax \citep{Navasardyan2008}, its
radio luminosity is low, and it may have re-brightened in radio at
$\sim 20$\,days, a sign of wind density variations \citep{Stockdale2008SN2008boRebrightening}.
However, its sparse radio data, do not determine its peak luminosity. Finally, SN\,2012P has an estimated hydrogen envelope mass of $<0.02\,M_\odot$ \citep{Fremling2016}, below the $\sim 0.1\,M_\odot$ threshold proposed by \citet{Chevalier_2010}, which suggests a compact progenitor. It was detected at 5\,GHz with a flux density of $0.44\pm0.05$\,mJy, $\sim 12$--$17$\,days after explosion \citep{Stockdale2012}. If this measurement lies near the 5\,GHz peak, the SSA model \citep{Chevalier_1998} gives a shell mean velocity typical of cIIb events. With a single radio epoch, however, neither the peak luminosity nor the peak time
can be determined, so the classification remains tentative. 

In the case of SN\,2025adtq, because of the lack of early, deep radio constraints, our observations provide only limited information on its possible association with the class of radio-faint SNe IIb (essentially because we cannot constrain the radio peak time). For example, if we assume that the SN\,2025adtq emission peaked around the time of our observations ($\sim 100-200$\,d after explosion), as in the case of SN\,2025ulz, then we can conclude that its ejecta were substantially slower than those of SN\,2025ulz. However, without a real constraint on the radio peak time, we cannot exclude that the radio emission of SN\,2025adtq peaked much earlier, at a luminosity comparable to that of the radio-faint SNe IIb discussed above. Finally, we argue that continued radio observations are valuable. Indeed, SN\,2025adtq could be a CSM-dominated event in which strong free--free absorption suppresses the emission at early times, as initially proposed for SN\,2010P \citep[see the rightmost black star in Figure~\ref{fig:Chevalier};][]{10.1093/mnras/stu430, Bietenholz2021RadioSNe}. Continued monitoring may reveal a late-time re-brightening that would support this interpretation.

\subsection{Jet afterglow constraints}

\begin{figure}
    \centering
    \includegraphics[width=0.5\textwidth]{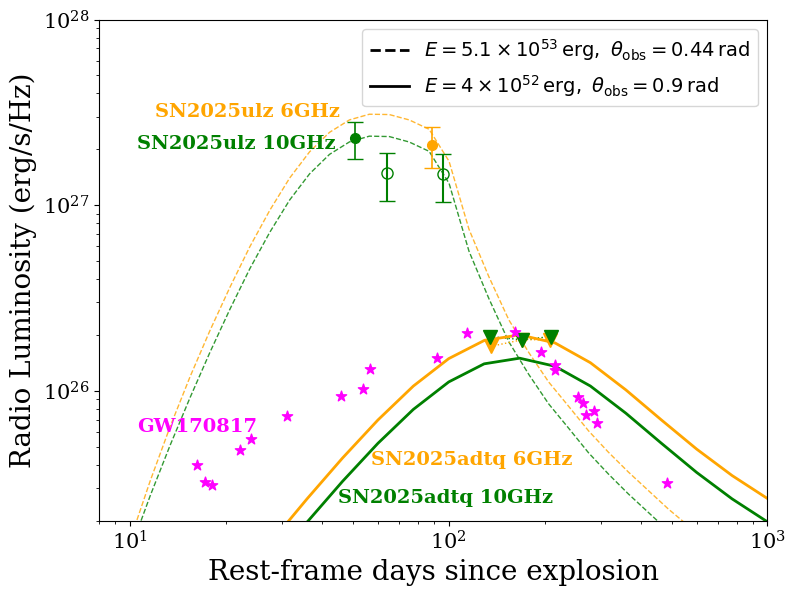}
    \caption{Radio observations and light curve models (lines, computed with \texttt{AfterglowPy:}\citealt{Ryan_2020}) of SN\,2025adtq (upper limits; downward triangles) and SN\,2025ulz \citep[circles;][]{O’Dwyer_2026}. The dashed lines represent the afterglow of a jet that may explain the radio counterpart of SN\,2025ulz. A jet with similar properties is clearly excluded in the case of SN\,2025adtq. However, we cannot exclude emission from a jet with lower isotropic energy, such as the one whose radio afterglow is plotted here with solid lines.  For comparison, we plot the $3$\,GHz light curve of GW170817 \citep[magenta stars;][]{Makhathini_2021}. 
    }
    \label{fig:Afterglow models}
\end{figure}

In Figure~\ref{fig:Afterglow models}, we compare the radio upper limits we collected for SN\,2025adtq with predictions for Gaussian jet afterglows computed using \texttt{afterglowpy} \citep{Ryan_2020}. As evident from this Figure, our deep radio limits exclude a jet similar to the one whose afterglow remains a viable explanation for the radio counterpart of SN\,2025ulz \citep[electron energy distribution index $p=2.133$; energy fractions $\epsilon_e=0.01$ and $\epsilon_B=0.001$; on-axis isotropic-equivalent kinetic energy $E_{\rm iso}=5.1\times10^{53}$\,erg; jet core angle $\theta_{\rm core}=4.01^\circ$; viewing angle $\theta_{\rm obs}=25.2^\circ$; and uniform ISM density $n_{\rm ISM}=3.1$\,cm$^{-3}$;][]{O’Dwyer_2026}. Indeed, at the luminosity distance of SN\,2025adtq ($\approx 389$\,Mpc), this model exceeds our radio upper limits (dashed lines in Figure~\ref{fig:Afterglow models}). This rules out the specific combination of jet and environmental parameters inferred for SN\,2025ulz in the case of SN\,2025adtq. However, as shown by the solid lines in Figure~\ref{fig:Afterglow models}, we cannot exclude jets with smaller isotropic-equivalent energies. Setting
$E_{\rm iso}=(4.0$--$5.3)\times10^{52}$\,erg, increasing the viewing angle to $\theta_{\rm obs}\approx48^\circ$--$55^\circ$, and adopting
$n_{\rm ISM}=1$\,cm$^{-3}$ yields models consistent with our upper limits, even if those limits had been obtained at the most constraining time, that is, at the peak of the radio emission.

\section{Summary and Conclusion}
\label{sec:conclusion}
S250818k and S251112cm were both identified as compact-binary merger candidates during the fourth observing run (O4) of the LVK detector network. Both may contain subsolar-mass components, and each has a proposed Type IIb SN counterpart. This makes a direct comparison of the two counterparts, presented in this work, worthwhile. S250818k was a low-significance GW trigger, with a reported false-alarm rate of 2.1\,yr$^{-1}$. Radio observations of its proposed counterpart, SN\,2025ulz, revealed faint but significant emission \citep{O’Dwyer_2026}. This emission is consistent with SN--CSM interaction, although an off-axis jet peaking $\sim$50--100\,days after the GW trigger also remains viable (Figure~\ref{fig:Afterglow models}).
S251112cm was a somewhat more significant GW trigger, with a reported false-alarm rate of $\sim 0.25$\,yr$^{-1}$ (about one per four years). Our radio observations of its proposed counterpart, SN\,2025adtq, did not detect any emission. The $3\sigma$ upper limits correspond to luminosities of
$L_\nu \lesssim 2\times10^{26}$\,erg\,s$^{-1}$\,Hz$^{-1}$ at the observed epochs. 

For SN\,2025ulz, the detected radio emission allowed us to interpret this event meaningfully in the context of both the SSA model and the off-axis jet model \citep{O’Dwyer_2026}. In the case of SN\,2025adtq, the lack of a detection limits our ability to constrain its properties in the context of the SSA model for radio SNe. Indeed, without a radio detection, the peak luminosity and peak time are unknown, so we cannot securely constrain the shock velocity of SN\,2025adtq or the mass-loss rate of its progenitor. Continued radio monitoring at later epochs could provide a valuable test of the CSM-interaction scenario.

Our radio limits are more informative about a relativistic outflow. For SN\,2025adtq, our non-detections rule out a jet with the parameters inferred for SN\,2025ulz. If we assume that we observed SN\,2025adtq around the time at which a hypothetical jet peaked in the radio band, then such a jet is constrained to be less energetic or viewed farther off-axis (Figure~\ref{fig:Afterglow models}). 

Although we can neither confirm nor rule out an association between S251112cm and SN\,2025adtq, the question of whether SNe coincident with subsolar-mass GW candidates can launch fast, collimated jets that produce bright radio emission remains compelling. The S250818k/SN\,2025ulz and S251112cm/SN\,2025adtq pairs illustrate how radio observations can address this question, through a detection in the first case and through constraining non-detections in the second. Future systematic follow-up of such candidates, including lower-significance GW triggers, should combine early radio observations with continued late-time monitoring, which together can best distinguish emission from non-relativistic SN--CSM interaction from that of relativistic, possibly off-axis, jets. Deep, late-time radio observations of SNe associated with subsolar-mass GW candidates could also provide an independent constraint on the amount of r-process material produced. As the r-process-rich ejecta decelerate in the surrounding medium, they are expected to produce late-time radio emission \citep{Nakar2011,Piran2013,Hotokezaka2015,Margalit2015,Hotokezaka2016,Kathirgamaraju2019}. Radio non-detections can therefore limit the ejecta mass \citep{2022ApJ...934L...5L}, complementing optical and near-infrared constraints on any kilonova emission, which can be difficult to disentangle from the SN light.

\begin{acknowledgements}
\small T.O. and A.C. acknowledge support from the National Science Foundation (NSF) via grant AST-2431072.  DK was suported by NSF grant AST-2511757. KPM acknowledges the Anusandhan National Research Foundation grant ANRF/ECRG/2024/005949. The National Radio Astronomy Observatory and Green Bank Observatory are facilities of the U.S. National Science Foundation operated under cooperative agreement by Associated Universities, Inc. We thank Avery Eddins for providing feedback on this work.
\end{acknowledgements}

\begin{table*}
\centering
\caption{VLA observations: SN name; MJD of the observation; rest frame days since S251112cm trigger (MJD=60991.64); observing frequency; flux density; VLA array configuration; FWHM of the VLA nominal synthesized beam; image RMS; VLA. All observations were conducted under project VLA/22B-235 PI: Corsi. 
\label{tab:Radio data}}

\begin{tabular}{cccccccc}
\toprule
SN & ${\rm T_{VLA}}$ &  ${\rm \Delta T_{\rm VLA}}$  & $\nu$  & $F_{\nu}$ & Conf. & Nom.Beam  & Image  RMS\\
&  (MJD) & (Day) & (GHz) & ($\mu$Jy) &  & (arcsec) & ($\mu$Jy)\\
\hline
2025adtq & 61131.03 & 139 & 10 & $<7.8$ & A & 0.20 & 2.6 \\
 & 61131.36 & 140 & 5.5 & $<6.9$ & A & 0.33 & 2.3 \\
 & 61166.97 & 175 & 10 & $<7.5$ & A & 0.20 & 2.5 \\
 & 61167.30 & 176 & 5.5 & $<7.5$ & A & 0.33 & 2.5 \\
 & 61207.81 & 216 & 10 & $<7.8$ & A & 0.20 & 2.6 \\
 & 61207.14 & 215 & 5.5 & $<7.5$ & A & 0.33 & 2.5 \\
\hline
2025adtq C-band & Combined & 139-216 & 5.5 & $<4.2$ & A & 0.33 & 1.4 \\
2025adtq X-band & Combined & 139-216 & 10 & $<4.8$ & A & 0.2 & 1.6 \\
\hline
\end{tabular}
\end{table*}

\bibliographystyle{aasjournal}
\bibliography{main}

\end{document}